\documentstyle[12pt]{article}
\def\be{\begin{equation}}
\def\ee{\end{equation}}
\def\bea{\begin{eqnarray}}
\def\eea{\end{eqnarray}}

\begin{document}

\begin{center}
{\Large{\bf Noncommutative  Superstring Worldsheet}}                  
										 
\vskip .5cm   
{\large Davoud  Kamani}
\vskip .1cm
 {\it Institute for Studies in Theoretical Physics and 
Mathematics (IPM)
\\  P.O.Box: 19395-5531, Tehran, Iran}\\
{\sl e-mail: kamani@theory.ipm.ac.ir}
\\
\end{center}

\begin{abstract} 

In this paper we consider the worldsheet of superstring as a noncommutative
space. Some additional terms can be added to the superstring action, such 
that for ordinary worldsheet they are zero. Expansion of this extended action
up to the first order of the noncommutativity parameter, leads to the new
supersymmetric action for string.
For the closed superstring, we obtain the boundary state that describes a 
brane. From the open string point of view, the new boundary conditions on
the worldsheet bosons, generalize the noncommutativity of spacetime.
Finally, we suggest some definitions for the  
noncommutativity parameter of superstring worldsheet.

\end{abstract} 



\newpage
\section{Introduction}

Noncommutative geometry plays a fascinating role in string theory.
There has been a great deal of recent interests in noncommutative theories,
stimulated by their connection with string theory and $M$-theory; for a
review and comprehensive list of references see Ref.\cite{1} and
\cite{2,3,4,5,6,7,8}.
The idea that the coordinates of spacetime do not commute at sufficiently
small distance scales is related to the non-perturbative backgrounds
of string and M theories \cite{2,3,4,5,8}. 
Noncommutativity on D-branes in the
presence of constant background $B$-field, was the original interest 
\cite{2,3,4}. The worldvolume of a D-brane with constant background $B$-field
is a simple and concrete example of a noncommutative spacetime, in which
 gauge and matter fields live \cite{2,6}.

Noncommutative field theories have rich structures. 
The embedding of these theories into
string theory \cite{3}, suggests that these structures may be directly
relevant to reconsidering the familiar notions of the superstrings and the
low energy limits of them. In other words, any change in the string theory
affects the whole noncommutativity. 
Now we introduce some of these changes.

We consider the worldsheet of superstring as a two dimensional
noncommutative space.
Therefore we can introduce some additional terms to the superstring action
that for the ordinary worldsheet they are zero. For the small  
noncommutativity parameter of the string worldsheet we develop the worldsheet 
supersymmetry for this action. The boundary conditions of open string with
noncommutative worldsheet, lead to the generalized noncommutativity 
parameter of spacetime. In this case the noncommutativity of 
spacetime is a consequence of $B$-field and
the noncommutativity of the string worldsheet. The closed string
emitted from a brane with background field, has a boundary state that
is generalized by the noncommutativity of its worldsheet. 

Our motivation for studying noncommutativity of string worldsheet is 
the following. If the worldsheet lives in a noncommutative spacetime, it is
natural to expect it to inherit the noncommutativity from the spacetime.
This can be seen from the fact that the pull-back of the spacetime
noncommutativity parameter on the string worldsheet is not zero.

It is worth emphasizing that such theory is inherently non-conformal.
The parameter of noncommutativity introduces a length scale in the
worldsheet which breaks the scale invariance and subsequently the 
conformal invariance of the theory. Despite lack of conformal invariance, 
for the following reasons we shall investigate the model.

From the renormalization group and flows, it is shown that large (small) 
distance of the spacetime corresponds to small (large) distance of
worldsheet. In other words we have the relation $L^2= ln(\Lambda/\mu)$,
where $\Lambda^{-1}$ is a characteristic two dimensional distance that
is very much shorter than the two dimensional distance $\mu^{-1}$ that
the worldsheet is seen and $L$ is a characteristic spacetime distance
\cite{9}. In fact $\Lambda$ is two dimensional UV cut-off.
Now consider finite UV cut-off. This will certainly break scale invariance
of the worldsheet theory. If we allow the scale invariance of the 
worldsheet to be broken at very short distances on the worldsheet, we
can interpret the worldsheet noncommutativity parameter as the UV cut-off
for the worldsheet.

This paper is organized as follows. In section 2, we briefly review 
superstring with ordinary worldsheet. In section 3, we present a
new action and corresponding supersymmetry for the superstring with 
noncommutative worldsheet. In section 4, we study closed string and
its boundary state, that describes a brane. In section 5, we obtain the
boundary conditions of open string, in presence of a brane. In section 6, 
some definitions for the noncommutativity  parameter of the worldsheet of
superstring is suggested. 
\section{Superstring with ordinary worldsheet}

Superstring in presence of a brane with background 
fields, is described by the action \cite{10}
\bea
S&=&-\frac{1}{4\pi \alpha'}\int_{\Sigma} d^2 \xi \bigg{(} \sqrt{-h}
h^{ab}g_{\mu \nu} \partial_a X^\mu \partial_b X^\nu + 
\epsilon^{ab} B_{\mu \nu} \partial_a X^\mu \partial_b X^\nu
-i\sqrt{-h}g_{\mu \nu} \bar{\psi}^\mu \rho^a \partial_a \psi^\nu 
\bigg{)}\nonumber\\
&~&+ \frac{1}{4\pi \alpha'}\int_{\partial \Sigma} d\zeta F_{\alpha \beta}
(X^\alpha \partial_\zeta X^\beta +\frac{i}{2} 
\theta^\alpha \theta^\beta)\;\;,
\eea
where, $\Sigma$ is the worldsheet of the string, and $\partial \Sigma$ is 
its boundary. 
The indices $\alpha, \beta,\gamma, ...$, show the brane directions.
Coordinate $\zeta$ is tangent to the boundary of the string
worldsheet. The field $B_{\mu \nu}$ is the NS$\otimes$NS massless field, and
$F_{\alpha \beta}$ is constant field strength of a $U(1)$ gauge field
$A_{\alpha}$. The field $\theta^\mu$ is the following combination of the 
components $\psi^{\mu}_1$ and $\psi^{\mu}_2$ of the worldsheet fermion
$\psi^\mu$,
\bea
\theta^\mu = \psi_1 ^{\mu}+i\psi_2^{\mu} \;\;.
\eea

Let $F_{\alpha \beta}=0$, $g_{\mu \nu}=\eta_{\mu \nu}
={\rm diag}(-1,1,...,1)$ and $B_{\mu \nu}$ be constant
background field. Also consider $h_{ab}=\eta_{ab}={\rm diag}(-1,1)$.
Therefore the equations of motion are,
\bea
&~&(\partial_{\tau}^2-\partial_{\sigma}^2)X^\mu = 0 \;\;,\nonumber\\
&~&\partial_+ \psi_1^\mu = 0\;\;,\nonumber\\
&~&\partial_- \psi_2^\mu = 0\;\;,
\eea
where, $\partial_{\pm}=\frac{1}{2}(\partial_{\tau} \pm \partial_{\sigma})$.
The invariance of the action under the worldsheet supersymmetry
transformations
\bea
&~& \delta X^\mu = \bar \epsilon \psi^\mu \;\;,\nonumber\\
&~& \delta \psi^\mu = -i\rho^a \partial_a X^\mu \epsilon \;\;,
\eea
leads to the following boundary state equations for the closed superstring
\bea
(\partial_{\tau} X^{\alpha} - B^\alpha_{\;\;\;\beta} 
\partial_{\sigma} X^{\beta}
)_{\tau_0} \mid B \rangle =0\;\;, 
\eea
\bea
(\partial_{\sigma}X^i)_{\tau_0} \mid B \rangle =0\;\;, 
\eea
for the bosonic part, and
\bea
\bigg{(} \psi_1 ^\alpha - i \psi_2 ^\alpha +B^\alpha_{\;\;\;\beta}
(\psi_1 ^\beta + i \psi_2 ^\beta)\bigg{)}_{\tau_0} \mid B \rangle =0\;\;, 
\eea
\bea
(\psi_1 ^i + i \psi_2 ^i)_{\tau_0} \mid B \rangle =0\;\;, 
\eea
for the fermionic part. 
The indices $i,j,...,$ show the transverse directions of the brane. 
Since the presence of the brane breaks half of the 
supersymmetry, for deriving (5)-(8) we used the relation $\epsilon_2 = 
i\epsilon_1$.

The boundary conditions of open superstring are
\bea
&~&(\partial_{\sigma} X^{\alpha} - B^\alpha_{\;\;\;\beta} 
\partial_{\tau} X^{\beta})_{\sigma_0} =0\;\;, 
\nonumber\\
&~&(\partial_{\tau}X^i)_{\sigma_0} =0\;\;, 
\nonumber\\
&~&\bigg{(} \psi_1 ^\alpha + i \psi_2 ^\alpha +B^\alpha_{\;\;\;\beta}
(\psi_1 ^\beta - i \psi_2 ^\beta)\bigg{)}_{\sigma_0} =0\;\;, 
\nonumber\\
&~&(\psi_1 ^i - i \psi_2 ^i)_{\sigma_0} =0\;\;, 
\eea
where the boundaries are at $\sigma_0 = 0,\pi$.
\section{Noncommutative worldsheet}

Let $\xi^0$ and $\xi^1$ be
the coordinates of the worldsheet of superstring. 
The ``star product'' between two arbitrary
functions $f(\xi^0 , \xi^1)$ and $g(\xi^0 , \xi^1)$ is 
\bea
f(\xi^0 , \xi^1) * g(\xi^0 , \xi^1) 
= \exp \bigg{(}\frac{i}{2} \theta^{ab} \frac{\partial}{
\partial \zeta^a}\frac{\partial}{\partial \eta^b} \bigg{)} 
f(\zeta^0 , \zeta^1) g(\eta^0 , \eta^1) \mid_{\zeta=\eta=\xi} \;\;.
\eea
Therefore there is noncommutativity between $\xi^0 $ and $\xi^1$, i.e.
\bea
\xi^a * \xi^b - \xi^b * \xi^a = i \theta^{ab}\;\;.
\eea
Later we shall discuss the antisymmetric tensor $\theta^{ab}$. 

For the coordinates $(\sigma , \tau)$ let $\eta_{ab}$ be
the metric of the string worldsheet,
therefore the superstring action under the star product becomes
\bea
S^* =-\frac{1}{4\pi \alpha'}\int_{\Sigma} d^2 \sigma \bigg{(} 
g_{\mu \nu} \partial_a X^\mu * \partial^a X^\nu + 
\epsilon^{ab} B_{\mu \nu} \partial_a X^\mu * \partial_b X^\nu
-ig_{\mu \nu} \bar{\psi}^\mu * \rho^a \partial_a \psi^\nu 
\bigg{)}+{\bar S^*}\;\;,
\eea
where $\epsilon^{01}=-\epsilon^{10}=1$.  
The action ${\bar S^*}$ contains the bosonic and the fermionic fields of 
the worldsheet, and when the star product changes to the usual product, i.e.
for $\theta^{ab}=0$, it vanishes.
We consider ${\bar S^*}$ as the following
\bea
{\bar S^*} = -\frac{1}{4\pi \alpha'}\int_{\Sigma} d^2 \sigma \bigg{(} 
C_{\mu \nu} X^\mu *  X^\nu + 
k^{ab} A_{\mu \nu} \partial_a X^\mu * \partial_b X^\nu
+\frac{i}{2}\epsilon^{ab}S_{\mu \nu} \bar{\psi}^\mu *
\rho_a \rho_b \psi^\nu \bigg{)}\;\;,
\eea
where $C_{\mu \nu}$ and $A_{\mu \nu}$ are arbitrary antisymmetric tensors and
$S_{\mu \nu}$ and $k^{ab}$ are arbitrary symmetric tensors. Many other terms
can be considered that their usual product vanish. For example the
terms
\bea
&~&C_{\mu \nu} \partial_{a_1}...\partial_{a_m} X^\mu *
\partial^{a_1}...\partial^{a_m} X^\nu \;\;, 
\nonumber\\
&~&k^{ab}A_{\mu \nu} \partial_{a_1}...\partial_{a_l}\partial_a X^\mu *
\partial^{a_1}...\partial^{a_l}\partial_b X^\nu \;\;,
\nonumber\\
&~&\epsilon^{ab}S_{\mu \nu} \partial_{a_1}...\partial_{a_n}\bar{\psi}^\mu *
\rho_a \rho_b \partial^{a_1}...\partial^{a_n}\psi^\nu \;\;,
\eea
are zero for the usual product. The arbitrary numbers $m,l$ 
and $n$ are positive integers. Because of the 
derivatives, we do not introduce these terms
to the action (13). After expanding in terms of the powers of $\theta^{ab}$,
the first non-zero term of the second term of the action (13) 
contains derivatives
of order four, for simplification this term is neglected too.
Also there are another terms such as $S_{\mu \nu} ^{(1)}\psi^\mu_1*
\psi_1^{\nu}$ and $S_{\mu \nu} ^{(2)}\psi_2^{\mu}*
\psi_2^{\nu}$ and their derivatives like (14), that  
for symmetric matrices $S_{\mu \nu}^{(1)}$ and $S_{\mu \nu} ^{(2)}$,
vanish under the usual product. 
These terms do not have worldsheet covariant forms,
therefore we also put away them.

Now we consider the expansion
of the action (12) up to the first order of $\theta^{ab}$ 
and study closed and open superstrings of it
\bea
S^*& =&-\frac{1}{4\pi \alpha'}\int_{\Sigma} d^2 \sigma \bigg{(} 
g_{\mu \nu} \partial_a X^\mu \partial^a X^\nu + 
\epsilon^{ab} B_{\mu \nu} \partial_a X^\mu \partial_b X^\nu
-ig_{\mu \nu} \bar{\psi}^\mu \rho^a \partial_a \psi^\nu 
\nonumber\\
&~&+\frac{1}{2}\theta^{ab}g_{\mu \nu} \partial_a {\bar \psi}
^\mu \rho^{a'} \partial_{a'} \partial_b \psi^\nu
-\frac{1}{4}\epsilon^{a'b'}\theta^{ab}S_{\mu \nu} \partial_a {\bar \psi}
^\mu \rho_{a'} \rho_{b'} \partial_b \psi^\nu
\nonumber\\
&~&+\frac{i}{2}\theta^{ab} C_{\mu \nu} \partial_a X^\mu \partial_b X^\nu
\bigg{)}+{\cal{O}}(\theta^2)\;.
\eea
The second term and the last three terms are total derivative.
Note that $\theta^{ab}$ has only one independent component, therefore it can
be written as
\bea
\theta^{ab}=\theta \epsilon^{ab}\;\;.
\eea
In the coordinate system $(\sigma , \tau)$ we choose $\theta$ as a 
constant parameter.

Let us define $B'_{\mu \nu}$ as follows 
\bea
B'_{\mu \nu}=B_{\mu \nu}+\frac{i}{2} \theta C_{\mu \nu}\;.
\eea
Therefore the $B$-term and $C$-term of the action (15)
can be combined to $B'$-term. If we
assume $C_{\mu \nu}$ to be a linear combination of $B_{\mu \nu}$
and $F_{\mu \nu}$ (field strength of a $U(1)$ gauge field)
\bea
C_{\mu \nu}=aF_{\mu \nu}+bB_{\mu \nu} \;\;,
\eea
gauge invariance of $B'_{\mu \nu}$ under the gauge transformations
\bea
&~&A_{\mu} \rightarrow A_{\mu}+\Lambda_{\mu}\;\;,
\nonumber\\
&~&B_{\mu \nu} \rightarrow B_{\mu \nu} + \partial_{\mu}\Lambda_{\nu} 
-\partial_{\nu}\Lambda_{\mu}\;\;,
\eea
requires the following relation between coefficients ``$a$'' and ``$b$''
and the parameter $\theta$
\bea
2+i\theta(a+b)=0\;\;.
\eea
This equation implies $a \neq -b$, which means if $B'_{\mu \nu}$ is a gauge 
invariant field, $C_{\mu \nu}$ in the form
of combination (18) is not gauge invariant.

From now on we neglect ${\cal{O}}(\theta^2)$ in the action (15). Let
$S_{\mu \nu}$ and $C_{\mu \nu}$ be constant, 
i.e. independent of the spacetime
coordinates. We introduce the new supersymmetry transformations,
\bea
&~& \delta X^\mu = \bar \epsilon \psi^\mu -i\theta S^\mu_{\;\;\;\nu} 
\partial_{\tau} ({\bar \epsilon}\psi^\nu) \;\;,
\nonumber\\
&~& \delta \psi^\mu = -i\rho^a \partial_a X^\mu \epsilon \;\;.
\eea
These transformations form a closed algebra. To see this, consider
two successive transformations with supersymmetry parameters $\epsilon$
and $\epsilon'$, therefore
\bea
[\delta_{\epsilon} , \delta_{\epsilon'}]X^\mu &=&
\delta_{\epsilon}(\delta_{\epsilon'}X^\mu)-(\epsilon \leftrightarrow
\epsilon')
\nonumber\\
&=&2i {\bar \epsilon} \rho^a \epsilon' (\partial_a X^\mu
-i\theta S^\mu_{\;\;\;\nu}\partial_{\tau}\partial_aX^\nu)\;\;,
\eea
for the worldsheet bosons, and
\bea
[\delta_{\epsilon} , \delta_{\epsilon'}]\psi^\mu =
2i {\bar \epsilon} \rho^a \epsilon' (\partial_a \psi^\mu
-i\theta S^\mu_{\;\;\;\nu}\partial_{\tau}\partial_a \psi^\nu)\;\;,
\eea
for the worldsheet fermions. To obtain the last equation, one should use
the equation of motion of $\psi^\mu$, which is $\rho^a \partial_a \psi^\mu
=0$.

As it is mentioned, the presence of a 
brane breaks half of the supersymmetry. 
For $\epsilon_2=i\epsilon_1 \equiv i{\varepsilon}$, the above
transformations become
\bea
&~&\delta X^\mu =-\varepsilon (\theta^\mu-i\theta S^\mu_{\;\;\;\nu}
\partial_{\tau}\theta^\nu) \;\;,
\nonumber\\
&~& \delta \psi_1 ^\mu = -2i\varepsilon \partial_- X^\mu \;\;,
\nonumber\\
&~& \delta \psi_2 ^\mu = 2\varepsilon \partial_+ X^\mu \;\;.
\eea
We shall use these transformations, to obtain the boundary conditions of
superstrings.
\section{Closed superstring}

For the closed superstring let the metric $g_{\mu \nu}$ be $\eta_{\mu \nu}$.
Now we concentrate to 
the R$\otimes$R and the NS$\otimes$NS 
sectors of type II superstring. These sectors
imply that the surface terms of the variation of the action (15) vanish.
This variation gives the boundary state equations for the closed 
superstring, emitted from the brane, as
\bea
(\partial_{\tau} X^{\alpha} - B'^\alpha_{\;\;\;\beta} 
\partial_{\sigma} X^{\beta}  - B'^\alpha_{\;\;\;\;i}
\partial_{\sigma} X^i 
)_{\tau_0} \mid B \rangle =0\;\;, 
\eea
\bea
(\delta X^i)_{\tau_0} \mid B \rangle =0\;\;, 
\eea
for the bosonic part. 
Equation (26) implies that $\partial_{\sigma}X^i$ vanishes on the
boundary, and will be dropped from the equation (25). 
From now on we assume that the 
mixed components of $S_{\mu \nu}$ are zero, i.e.
\bea
S_{i\alpha}=0\;\;.
\eea
According to the supersymmetry transformations (24) and the bosonic part of
the boundary state equations, i.e. equations (25) and (26), there are the
following boundary state equations for the worldsheet fermions
\bea
\bigg{(}\psi_1 ^i + i \psi_2 ^i -i \theta S^i_{\;\;j}\partial_{\tau}
(\psi_1 ^j + i \psi_2 ^j)
\bigg{)}_{\tau_0} \mid B \rangle =0\;\;, 
\eea
\bea
&~&\bigg{(} \psi_1 ^\alpha - i \psi_2 ^\alpha+B'^\alpha_{\;\;\;\beta}
(\psi_1 ^\beta + i \psi_2 ^\beta)+i\theta 
S^\alpha_{\;\;\;\beta}\partial_{\tau}
(\psi_1 ^\beta - i \psi_2 ^\beta)\nonumber\\
&~&-i\theta B'^\alpha_{\;\;\;\beta}S^\beta_{\;\;\;\gamma}\partial_{\tau}
(\psi_1 ^\gamma + i \psi_2 ^\gamma) 
\bigg{)}_{\tau_0} \mid B \rangle =0\;\;.      
\eea
As expected these equations respect the supersymmetry transformations.
We explicitly show this. That is, from the fermionic boundary conditions 
(28) and (29) and the supersymmetry transformations, we obtain the bosonic
boundary conditions (25) and (26). The equation (28) and the first
transformation of (24) give the transverse bosonic boundary condition (26).

To see  the consistency of (25) and (29), let us write the supersymmetry
transformations of the left and the right moving parts of $X^\mu$
\bea
&~& \delta X^\mu_L = -i\varepsilon (\psi^\mu_2 -\theta S^\mu_{\;\;\nu} 
\partial_{\tau}\psi^\nu_1) \;\;,
\nonumber\\
&~& \delta X^\mu_R = -\varepsilon (\psi^\mu_1 +\theta S^\mu_{\;\;\nu} 
\partial_{\tau}\psi^\nu_2) \;\;.
\eea
The sum of these transformations gives $\delta X^\mu $ of (24). 
The difference of these gives
\bea
\delta X'^\mu &=& \delta X^\mu_L-\delta X^\mu_R 
\nonumber\\
&=& \varepsilon (\lambda^\mu +i\theta S^\mu_{\;\;\nu} 
\partial_{\tau}\lambda^\nu) \;\;.
\eea
where $\lambda^\mu$ is
\bea
\lambda^\mu \equiv \psi^\mu_1 -i\psi^\mu_2 \;\;.
\eea
From the equation (29) we have
\bea
\bigg{(} \varepsilon(\lambda^\alpha+i\theta S^\alpha_{\;\;\;\beta}
\partial_{\tau} \lambda^{\beta})-B'^\alpha_{\;\;\;\beta}
[-\varepsilon(\theta^\beta-i\theta S^\beta_{\;\;\;\gamma}
\partial_{\tau} \theta^{\gamma})]\bigg{)}
_{\tau_0} \mid B \rangle =0\;\;.      
\eea
According to the transformations (24) and (31) this is
\bea
(\delta X'^\alpha - B'^\alpha_{\;\;\;\beta}\delta X^\beta)
_{\tau_0} \mid B \rangle =0\;\;,      
\eea
which is equivalent to the equation
\bea
(\partial_\sigma X'^\alpha - B'^\alpha_{\;\;\;\beta}
\partial_\sigma X^\beta)_{\tau_0} \mid B \rangle =0\;\;.      
\eea
For the coordinate $X'^\mu$ we have the relation
\bea
\partial_\sigma X'^\mu=\partial_\tau X^\mu\;\;,
\eea
that can be seen from the solution of the equation of motion,
\bea
X^\mu_L=x^\mu_L+2\alpha'p^\mu_L(\tau+\sigma)+\frac{i}{2}\sqrt{2\alpha'}
\sum_{n \neq 0}\frac{1}{n}{\tilde \alpha}^\mu_n e^{-2in(\tau+\sigma)}\;\;,
\nonumber\\
X^\mu_R=x^\mu_R+2\alpha'p^\mu_R(\tau-\sigma)+\frac{i}{2}\sqrt{2\alpha'}
\sum_{n \neq 0}\frac{1}{n}\alpha^\mu_n e^{-2in(\tau-\sigma)}\;\;,
\eea
where $X^\mu=X^\mu_L+X^\mu_R$ and $X'^\mu=X^\mu_L-X^\mu_R$. Therefore 
equations (35) and (36) give the boundary state equation (25).
Now we obtain the boundary state $\mid B \rangle$.

{\bf Boundary state}

Combining the solutions of the equations 
of motion and boundary state equations, 
we obtain these equations in terms of modes. Consider some
of the brane directions and some of the transverse directions of the brane
to be compact on tori. 

The boundary state of the bosonic part is 
\bea
\mid B_{bos} , \tau_0 \rangle = \sum_{\{p^\alpha\}}
\mid B_{bos} , \tau_0 , p^\alpha \rangle \;\;,
\eea
where
\bea
\mid B_{bos} , \tau_0 ,p^\alpha \rangle
&=&\frac{T_p}{2}\sqrt{\det(1+B')}\exp{\bigg{(}i\alpha'\tau_0 \sum_i(p^i_{op})
^2\bigg{)}}
\nonumber\\
&~&\times\delta^{(9-p)}(x^i-y^i)\exp \bigg{(} - \sum_{m=1}^{\infty}
\frac{1}{m}e^{4im\tau_0}\alpha^\mu_{-m} \Phi_{\mu \nu}
{\tilde \alpha}^{\nu}_{-m} \bigg{)}
\nonumber\\
&~&\times \mid 0 \rangle \prod_i \mid p^i_L=p^i_R=0 \rangle \prod_{\alpha}
\mid p^\alpha \rangle \;.
\eea
The set $\{y^i\}$ shows the position of the brane. The orthogonal matrix
$\Phi^\mu_{\;\;\;\nu}$ is
\bea
&~&\Phi^\mu_{\;\;\;\nu}=
(Q^\alpha_{\;\;\;\beta}\;,\;-\delta^i_{\;\;\;j})\;\;,
\nonumber\\
&~& Q^\alpha_{\;\;\;\beta}=[(1+B')^{-1}(1-B')]^\alpha_{\;\;\;\beta}\;\;\;.
\eea
The state (39) is general form of the state of Ref.\cite{11}.
The momentum of the closed string along the compact directions of the
brane, i.e. $\{X^{\alpha_c}\}$, is
\bea
&~&p^{\alpha_c}=\frac{1}{\alpha'}B'^{\alpha_c}
_{\;\;\;{\beta_c}}L^{\beta_c}\;\;,
\nonumber\\
&~&L^{\beta_c} = N^{\beta_c}R^{\beta_c}\;\;\;,
\eea
where $R^{\beta_c}$ is the 
radius of compactification of $X^{\beta_c}$-direction  
and $N^{\beta_c}$ is winding number of closed string around the   
$X^{\beta_c}$-direction. For interpretation of (41) see Ref.\cite{11}.  

For the NS$\otimes$NS sector, we have the following
fermionic boundary state equations
\bea
\bigg{(} (1-2r\theta S)^i_{\;\;\;j}\;b^j_re^{-2ir\tau_0}+
i(1+2r\theta S)^i_{\;\;\;j}\;{\tilde b}^j_{-r}e^{2ir\tau_0}
\bigg{)} \mid B_f , \tau_0 \rangle_{NS} =0\;\;,
\eea
for the transverse directions of the brane. For the directions along the 
brane we have
\bea
&~&\bigg{(}[1+B'+2r\theta(1-B')S]\;
^\alpha_{\;\;\;\beta}\;b^\beta_re^{-2ir\tau_0}
\nonumber\\
&~&-i[1-B'-2r\theta(1+B')S]^\alpha_{\;\;\;\beta}\;{\tilde b}
^\beta_{-r}e^{2ir\tau_0}
\bigg{)} \mid B_f , \tau_0 \rangle_{NS} =0\;.
\eea
In both of these equations, ``$r$'' is negative or positive 
half-integer number.

Equations (42) and (43) have the following solution
\bea
\mid B_f , \eta ,\tau_0 \rangle_{NS}
&=&K_{NS}\exp\bigg{[}i\eta \sum_{r=1/2}^{\infty}
\bigg{(}e^{4ir\tau_0}b^\mu_{-r} \Phi^{(r)}_{\mu \nu}
{\tilde b}^{\nu}_{-r} \bigg{)}\bigg{]}
\mid 0 \rangle \;\;,
\eea
where $\eta=\pm 1$ is introduced to make GSO projection easily. The matrix
$\Phi^{(r)}_{\mu \nu}$ has definition
\bea
\Phi^\mu_{(r)\;\nu} = (\Lambda^\alpha_{(r)\;\beta} \;,\; -H^i_{(r)\;j})
\;\;\;,\;\;\;\;\; r \geq \frac{1}{2}
\eea
\bea
H^i_{(r)\;j}=[(1-2r\theta S)^{-1}(1+2r \theta S)]^i_{\;\;\;\;j}\;\;,
\eea
\bea
\Lambda^\alpha_{(r)\;\beta}
=\bigg{(}[1+B'+2r\theta(1-B')S]^{-1}
[1-B'-2r\theta(1+B')S] \bigg{)}^\alpha_{\;\;\;\;\beta}\;\;.
\eea

Consistency of the solutions of equation (43) for positive and negative
``$r$'' requires the following relation between $B'^\alpha_{\;\;\;\beta}$
and $S^\alpha_{\;\;\;\beta}$, 
\bea
B'^\alpha_{\;\;\;\beta}(S^2)^\beta_{\;\;\;\gamma}
=(S^2)^\alpha_{\;\;\;\beta} B'^\beta_{\;\;\;\;\gamma}\;\;.
\eea
That is, $B'$ and $S^2$ should commute.
This is a restriction that naturally arises on $C$ and $S$.

The factor $K_{NS}$ is expected by the path integral with boundary action
\bea
K_{NS}=\prod_{r=1/2}^{\infty} \bigg{(} \det [1+B'+2r\theta 
(1-B')S]^\alpha_{\;\;\;\;\beta}\bigg{)}\;\;.
\eea
This is general form of the result \cite{12}.
For the ordinary worldsheet i.e. $\theta=0$, 
this reduces to ``1'' , as expected
(note that, $\sum_{r=1/2}^{\infty} 1 \leftrightarrow
{\rm lim}_{t \rightarrow 0} (2^t-1) \zeta(t)=0$ ).
The assumption of smallness of $\theta$, gives
\bea
K_{NS} = 1+\frac{\theta}{24} {\rm Tr}[(Q_0 S)^\alpha_{\;\;\;\beta}]+
{\cal{O}}(\theta^2)\;\;,
\eea
where $Q_0$ is given by (40) for $\theta=0$. Note that we made use of
$\sum_{r=1/2}^\infty r \leftrightarrow -\frac{1}{2}\zeta(-1)=\frac{1}{24}$ ,
and
\bea
\det (1+\theta M)=1+\theta {\rm Tr}(M) +{\cal{O}}(\theta^2)\;\;,
\eea
for a matrix $M$ to obtain (50). 
Up to the first order of $\theta$, $C$ does not appear
in $K_{NS}$.

For the R$\otimes$R sector, the boundary state equations of the worldsheet
fermions in terms of the modes are
\bea
(d^i_0+i{\tilde d}^i_0) \mid B_f , \tau_0 \rangle_R=0\;\;,
\eea
\bea
\bigg{(} (1-2n\theta S)^i_{\;\;\;j}\;d^j_n e^{-2in\tau_0}+
i(1+2n\theta S)^i_{\;\;\;j}\;{\tilde d}^j_{-n}e^{2in\tau_0}
\bigg{)} \mid B_f , \tau_0 \rangle_R =0\;\;,          
\eea
for the transverse directions of the brane, and
\bea
(d^\alpha_0-iQ^\alpha_{\;\;\;\beta}{\tilde d}^\beta_0)
 \mid B_f , \tau_0 \rangle_R =0\;\;,          
\eea
\bea
&~&\bigg{(}[1+B'+2n\theta(1-B')S]^\alpha_{\;\;\;\beta}
\;d^\beta_n e^{-2in\tau_0}
\nonumber\\
&~&-i[1-B'-2n\theta(1+B')S]^\alpha_{\;\;\;\beta}\;{\tilde d}
^\beta_{-n}e^{2in\tau_0}
\bigg{)} \mid B_f , \tau_0 \rangle_R =0\;,
\eea
for the brane directions. In the equations (53) and (55) 
the number ``$n$'' is a non-zero integer.

The solution of the equations (52)-(55) is
\bea
\mid B_f , \eta ,\tau_0 \rangle_R
&=&K_R\exp \bigg{[}i\eta \sum_{n=1}^{\infty}
\bigg{(}e^{4in\tau_0}d^\mu_{-n} \Phi^{(n)}_{\mu \nu}
{\tilde d}^{\nu}_{-n} \bigg{)}\bigg{]}
\mid B_f,\eta \rangle^{(0)}_R \;\;,
\eea
where $\mid B_f,\eta \rangle^{(0)}_R $ is solution of equations (52)
and (54), \cite{13,14}
\bea
\mid B_f,\eta \rangle^{(0)}_R ={\cal{M}}^{(\eta)}_{AB} \mid A \rangle
\mid {\tilde B} \rangle \;\;,
\eea
where $\mid A \rangle$ and $\mid {\tilde B} \rangle$
describe the vacuum of the fermionic zero modes $d^\mu_0$ and
${\tilde d}^{\mu}_0$. The matrix ${\cal{M}}^{(\eta)}$ is \cite{13, 15},
\bea
{\cal{M}}^{(\eta)}={\bar C}\Gamma^0\Gamma^{{\bar \alpha}_1}...
\Gamma^{{\bar \alpha}_p}\bigg{(} \frac{1+i\eta \Gamma_{11}}{1+i\eta}
\bigg{)}\exp(-\frac{1}{2}B'_{\alpha \beta} \Gamma^\alpha \Gamma^\beta)\;\;,
\eea
where ``${\bar C}$'' is 
charge conjugation matrix. Also brane is along the directions
$\{X^{{\bar \alpha}_1} ,...,X^{{\bar \alpha}_p}\}$.
Note that for the exponential in (58) there is a convention:
the exponential must be expanded, with the convention that all gamma
matrices anticommute, therefore there are a finite number of terms.

Again consistency of the solutions of equation (55), for positive and
negative ``$n$'' leads to the condition (48).

For the R$\otimes$R sector of superstring the matrices
$\Phi_{(n)}$, $H_{(n)}$ and $\Lambda_{(n)}$ are
\bea
\Phi^\mu_{(n)\;\nu} = (\Lambda^\alpha_{(n)\;\beta} \;,\; -H^i_{(n)\;j})
\;\;\;,\;\;\;\;\; n \geq 1
\eea
\bea
H^i_{(n)\;j}=[(1-2n\theta S)^{-1}(1+2n \theta S)]^i_{\;\;\;j}\;\;,
\eea
\bea
\Lambda^\alpha_{(n)\;\beta}
=\bigg{(}[1+B'+2n\theta(1-B')S]^{-1}
[1-B'-2n\theta(1+B')S] \bigg{)}^\alpha_{\;\;\;\;\beta}\;\;.
\eea
The factor $K_R$ is
\bea
K_R=\prod_{n=1}^{\infty} \bigg{(} \det [1+B'+2n\theta 
(1-B')S]^\alpha_{\;\;\beta}\bigg{)}\;\;\;.
\eea
For the ordinary worldsheet, this factor reduces to the expected result 
$\bigg{(} \det[(1+B)^\alpha_{\;\;\;\beta}] \bigg{)}^{-1/2}$ of Ref.\cite{12},
(note that, $\sum_{n=1}^{\infty}1 \leftrightarrow \zeta(0)=-\frac{1}{2}$).
The parameter $\theta$ is small, therefore
\bea
K_R = \frac{1}{\sqrt{\det[(1+B)^\alpha_{\;\;\;\beta}]}}\bigg{[} 1-\frac
{\theta}{2} \bigg{(} \frac{i}{2} {\rm Tr}[(1+B)^{-1}C]^\alpha_{\;\;\;\beta}
+\frac{1}{3}{\rm Tr}[(Q_0 S)^\alpha_{\;\;\;\beta}]\bigg{)}\bigg{]}
+{\cal{O}}(\theta^2)\;\;,
\eea
where we have used  
$\sum_{n=1}^\infty n \leftrightarrow \zeta(-1)=-\frac{1}{12}$ .

\section{Open superstring}

Now we obtain the boundary conditions of open superstring. From now on
consider the metric of the spacetime to be constant $g_{\mu \nu}$ . 
Also let  the mixed components of the metric be zero,
i.e. $g_{\alpha j}=0$. Furthermore assume that the field $B'$ has
non-zero components only along the brane, i.e. the components
$B'_{ij}$ and $B'_{\alpha j}$ are zero. 
The variation of the action (15) gives the boundary conditions
\bea
(\delta X^i)_{\sigma_0} =0\;\;, 
\eea
\bea
(g_{\alpha \beta}\partial_{\sigma} X^{\beta} - B'_{\alpha \beta} 
\partial_{\tau} X^{\beta}
)_{\sigma_0} =0\;\;, 
\eea
for the bosonic part, where $\sigma_0 =0, \pi$ show the boundaries. 
The worldsheet fermions obey the following boundary conditions
\bea
\bigg{(}g_{ij}(\psi_1 ^j- i \psi_2 ^j) -i \theta S_{ij}\partial_{\tau}
(\psi_1 ^j - i \psi_2 ^j)
\bigg{)}_{\sigma_0} =0\;\;, 
\eea
\bea
&~&\bigg{(} g_{\alpha \beta}(\psi_1 ^\beta + 
i \psi_2 ^\beta)+B'_{\alpha \beta}
(\psi_1 ^\beta - i \psi_2 ^\beta)+i\theta S_{\alpha \beta}\partial_{\tau}
(\psi_1 ^\beta + i \psi_2 ^\beta) 
\nonumber\\
&~&-i\theta B'_{\alpha \beta}S^\beta_{\;\;\;\gamma}\partial_{\tau}
(\psi_1 ^\gamma - i \psi_2 ^\gamma) 
\bigg{)}_{\sigma_0} =0\;.      
\eea
Similar to the closed superstring, one can show that these boundary 
conditions respect the worldsheet supersymmetry.
The open string boundary conditions (64)-(67), can be obtained from the
closed one, with the exchanges $\partial_{\tau} X^\mu\leftrightarrow
\partial_{\sigma}X^\mu$ and $\psi_1^\mu \rightarrow-\psi_1^\mu$.
This is equivalent to the change $\epsilon_2 \rightarrow -\epsilon_2$,
in supersymmetry transformations (21).

According to the boundary condition (64), the transverse directions of the
brane remain ordinary. Boundary condition (65) says that 
the worldvolume of the brane is a noncommutative space.
The parameter of spacetime noncommutativity is \cite{2}
\bea
\Theta^{\mu \nu}=-2\pi \alpha' \bigg{(}\frac{1}{g+B'}B'\frac{1}{g-B'}
\bigg{)}^{\mu \nu}\;\;.
\eea
The appearance of $B'$ instead of $B$ in this 
quantity shows the effects of the 
noncommutativity of the worldsheet to the spacetime noncommutativity. 
Thus for 
non-zero ``$\theta$'' and ``$C$'', the brane directions are noncommutative,
even if $B$-field vanishes.

If we apply the assumption of the smallness of $\theta$ in (68), we obtain
\bea
\Theta^{\mu \nu}=\Theta^{\mu \nu}_0 + \frac{i}{2}\theta \Omega^{\mu \nu}
+{\cal{O}}(\theta^2)\;,
\eea
where the matrix $\Omega$ is
\bea
\Omega= \Theta_0C(g-B)^{-1}-(g+B)^{-1}C \Theta_0
-2\pi \alpha'(g+B)^{-1}C(g-B)^{-1}\;,
\eea
as expected, $\Omega$ is an antisymmetric matrix.
The parameters $\Theta^{\mu \nu}_0$ show the spacetime noncommutativity 
for the ordinary string worldsheet. 

The effective metric of the open string is \cite{2}
\bea
&~&G_{\mu \nu}=g_{\mu \nu}-(B'g^{-1}B')_{\mu \nu}
= G^{(0)}_{\mu \nu}-\frac{i}{2}\theta (Bg^{-1}C+Cg^{-1}B)_{\mu \nu}
+\frac{1}{4}\theta^2 (Cg^{-1}C)_{\mu \nu} \;\;,
\nonumber\\
&~&G^{\mu \nu}=\bigg{(}\frac{1}{g+B'}g\frac{1}{g-B'}
\bigg{)}^{\mu \nu}\;\;.
\eea
Up to the order $\theta$, $G^{\mu \nu}$ is
\bea
G^{\mu \nu}=
G_0^{\mu \nu}+\frac{i}{2} \theta \bigg{(}G_0C(g-B)^{-1}-(g+B)^{-1}
CG_0\bigg{)}^{\mu \nu}+{\cal{O}}(\theta^2)\;\;,
\eea
where $G^{\mu \nu}_0$ and $G^{(0)}_{\mu \nu}$ refer to
the metric that is seen by open string with ordinary worldsheet.

Now we use the metric (71) to calculate the first 
correction of Yang-Mills and open string couplings
\bea
\frac{1}{g^2_{YM}}=\frac{(\alpha')^{(3-p)/2}}{(2\pi)^{p-2}g_s}\bigg{(}
\frac{\det(g+B')}{\det G} \bigg{)} ^{1/2} \;\;,
\eea
\bea
G_s= \frac{(\alpha')^{(3-p)/2}}{(2\pi)^{p-2}}g^2_{YM}\;.
\eea
These give
\bea
g_{YM}=g^{(0)}_{YM}\bigg{(}1+\frac{i}{8}\theta {\rm Tr}[(g+B)^{-1}C
]\bigg{)}+{\cal{O}}(\theta^2)\;,
\eea
\bea
G_s=G^{(0)}_s \bigg{(}1+\frac{i}{4}\theta {\rm Tr}[(g+B)^{-1}C
]\bigg{)}+{\cal{O}}(\theta^2)\;,
\eea
where $g^{(0)}_{YM}$ and $G^{(0)}_s $ are the Yang-Mills and open string 
couplings for the ordinary worldsheet, in noncommutative spacetime.
\section{The parameter of the noncommutativity}
Now we suggest some definitions for the noncommutativity parameter of 
the string worldsheet. These definitions 
are independent of the assumption of 
the smallness of $\theta^{ab}$, that we used in previous sections.
If we change the worldsheet coordinates $\xi^0$ and $\xi^1$
to $\xi'^0=\xi'^0 (\xi^0 , \xi^1)$ and $\xi'^1=
\xi'^1 (\xi^0 , \xi^1)$, the tensor
$\theta^{ab}$ changes to $\theta'^{a'b'}$, 
\bea
\theta'^{a'b'}=\frac{\partial \xi'^{a'}}{\partial \xi^a}
\frac{\partial \xi'^{b'}}{\partial \xi^b}\theta^{ab}\;\;.
\eea
As expected, this implies that noncommutativity depends on 
the coordinate system of the string worldsheet.
Note that according to the relation (77) we can choose a 
coordinate system on the string worldsheet with constant noncommutativity
parameter, i.e. independent  of the worldsheet coordinates.

{\bf The first definition}

Since the quantity $\theta_{ab}\theta^{ab}$ 
does not change from one coordinate system of 
worldsheet to another one we give the first definition of $\theta^{ab}$ as
\bea
\theta_{ab} \theta^{ab}=\Theta_{\mu \nu}\Theta^{\mu \nu}\;\;.
\eea
In the coordinate system $(\sigma ,\tau)$, the 
left hand side is $-2 \theta^2$. 
Raising the indices of $\Theta_{\mu \nu}$ leads to
\bea
\theta^2= \frac{1}{2} \Theta^{\mu \nu} G_{\nu \nu'} 
\Theta^{\nu' \mu'} G_{\mu' \mu}
\equiv \frac{1}{2} {\rm Tr}(\Theta G \Theta G)\;\;.
\eea
Again for $C_{\mu \nu} \neq 0$, the right hand side also contains 
$\theta$. Therefore (79) is an equation for $\theta$. 

{\bf The second definition}

Consider the noncommutative Yang-Mills theory and background dependence 
of it \cite{2}. For the background $B$, the noncommutativity
of spacetime is described by $\Theta_0$, and for $B'$, it is
described by $\Theta$. It has been discussed in \cite{2} 
that, the background independence
of noncommutative Yang-Mills at fixed ``$g_{\mu \nu}$'', leads to this
fact that, the quantity $g^2_{YM} \sqrt{\det \Theta}$ must be invariant 
under the changes of the background field. Therefore
\bea
g^2_{YM} \sqrt{\det \Theta}=(g^{(0)}_{YM})^2 \sqrt{\det \Theta_0}\;\;.
\eea
We suggest this equation as second definition for the parameter $\theta$.
According to the equations (68) and (73) the left hand side is a function 
of $\theta$, therefore from this equation $\theta$ is calculated. 
Note that for zero slope limit \cite{2}, the above equation 
reduces to an identity, i.e. the left hand side will be independent of
$\theta$.
\section{Conclusions and remarks}
The noncommutative worldsheet of superstring affects many  things.
The additional terms to the noncommutative action of superstring, 
generalize the supersymmetry transformations,
the boundary state of closed superstring, the boundary conditions of open 
superstring, Yang-Mills and open string couplings, and many other things.
The new closed string boundary state describes a brane that is more 
general than the mixed branes \cite{11,15}. 
The noncommutativity of the string worldsheet also changes the spacetime
noncommutativity. Therefore even if the background 
$B$-field vanishes, spacetime remains noncommutative.

We suggested some definitions for the noncommutativity parameter 
of the string worldsheet that relate this parameter to the corresponding 
one of spacetime.

As it has been discussed in Ref.\cite{9}, renormalization exhibits
the large distance spacetime physics to be encoded in the short distance
structure of the worldsheet. In other words the renormalization is justified
by the divergence of $L^2= ln(\Lambda/\mu)$. 

According to the renormalization group, the short two dimensional UV
cut-off distance $\Lambda^{-1}$ slides to more and more short distance
$\Lambda_0^{-1}$. In other words, there is an effective worldsheet at two
dimensional distance $\Lambda^{-1}$.
The effective worldsheet can be used to calculate effective string effects
at spacetime distance $L$. Since the noncommutativity parameter of the
worldsheet breaks the scale invariance of the worldsheet theory, it can
be interpreted as the UV cut-off. Therefore the cut-off distances
$\Lambda_0^{-1}$ and $\Lambda^{-1}$ correspond to two noncommutativity
parameters $\theta_0$ and $\theta$ respectively. This implies that for the 
noncommutativity parameter of the worldsheet, there are some bounds. 

{\bf Acknowledgements} 

I would like to thank H. Arfaei for useful discussion. Also I am grateful
to referee of the ``European Physical Journal C'', for useful comments,
which help me to improve the manuscript. 

\end{document}